The study of an interaction between the jet and an interstellar medium around knot E and knot F of radio galaxy M87 by *Chandra*

S.Osone

Chiba, Japan
osonesatoko@gmail.com

Abstract
The jet may compress an interstellar medium and soft X ray outside the jet may be absorbed by a compressed interstellar medium and thermal emission heated by a shock is expected inside the jet. I analyze an X ray image of the surrounding of the jet of M87 by *Chandra*. There is an original work for an image analysis. In this time, I select data for an image analysis by a removal of a pile up event completely and a precession consideration. I confirm a dip in soft X ray outside the jet between knot E and knot F with 224 ks archival data. I analyze X ray energy spectra for knot E and knot F with 480 ks archival data by *Chandra* in order to search for thermal emission. However, X ray energy spectra for knot E and knot F are well described with synchrotron emission.

1.Introduction
M87 is a radio galaxy and located in the center of Virgo cluster. The distance is 16.8 Mpc (Blakeslee et al. 2009). M87 is famous for Blackhole shadow observation by EHT (EHT 2019).

M87 is observed from a radio to TeV gamma ray. M87 have an inclined jet with a length of 20”. The angular resolution is micro second in a radio, 0”.7 in an optical band and 0”.5 in an X ray by *Chandra*, 5’.2 in GeV gamma ray by *Fermi*, 2’ in TeV gamma ray by air Cherenkov detector as CTA, HESS, LHASSO, MAGIC and VERITAS, 0.1 degree in TeV gamma ray by water Cherenkov detector as HWAC and LHASSO, and 0.5 degree in TeV gamma ray by air shower arrays as LHASSO and Tibet AS$\gamma$. Therefore, the jet cannot be resolved in high energy gamma ray. However, the origin of high energy emission can be limited by a timescale of flux variability and by simultaneous observation with multiwave length. A time scale of flux variability is 2d in TeV gamma ray (Aharonian et al. 2006). This means that the size of emitting area is $\delta c t_{val} = 3.1\times10^{16} (\delta/6)$ cm. Here, c is a speed of light and δ is a beaming factor. Therefore, the origin of TeV gamma ray is considered as nucleus. TeV flare occurred in 2018 and 2022 since 2010. Algaba et al. (2024) and Cao et al.(2024) also showed nucleus as the origin of TeV gamma ray from flux variability in TeV gamma ray. A flux correlation between nucleus in radio band and TeV gamma ray in 2008 showed nucleus as origin of TeV gamma ray (Acciari et al. 2009). A flux correlation between nucleus in X ray and TeV gamma ray in 2008 and 2010, between HST-1 in X ray and TeV gamma ray in 2005 are reported (Abramowaski et al. 2012). Therefore, HST-1 is also considered as the origin of TeV gamma ray.

The standard interpretation of multiwave length energy spectra for the M87 jet is synchrotron inverse compton model of accelerated electrons. The energy spectra is described by synchrotron emission in a low energy and inverse compton model in a high energy. The identical origin of electrons expects same flux variability in all energy band. Benkhali et al(2019) analyze *Fermi* data from 2008 to 2016 and show a chance probability of non flux variability is 0.018 below 10 GeV and 0.23 above 10 GeV. Cao et al.(2024) analyze *Fermi* data from 2009 to 2024 and show non flux variability with three months bin, but show flux variability with a short time scale bin of day. Algaba et al.(2024) also analyzed *Fermi* data in 2018 and reported flux variability with bin of few days. However, count per short time scale is poor statistics and significance of flux variability is not reported. Flux variability in GeV gamma ray is debated yet. There are some models which show non flux variability in GeV gamma ray. One is a pion decay. as an interaction between accelerated protons and an interstellar medium. This energy spectra peaks at 80 MeV, which is a half energy of pion. However, energy spectra of M87 have not been reported below 1 GeV by *Fermi*. Another is non thermal bremsstrahlung as an interaction between accelerated electrons and an interstellar medium. Osone(2017) show no contribution of non thermal bremsstrahlung from both nucleus and HST-1 to observed flux with *Fermi*.

Dainotti et al(2012) made a merged image with archival data from 2000 to 2009 by *Chandra* and reported that there is a soft X ray dip outside the jet between knot E and knot F as the interaction between the jet and an interstellar medium. There is a pile up problem for CCD of *Chandra*. A pile up distort both an energy spectra and an image. For a merged image, they removed data with a visible pile up line originated from nucleus or HST-1. However, they did not remove data with pile up event in a count rate. For correct image analysis, pile up event have to be removed completely. There are two kinds of apprant movements. One is a jet apprant velocity, the other is a precession. When an image is merged, these apprant movements have to be considered.

I use plenty of archive data with *Chandra* and make a merged image by a removal of pile up event completely and a consideration of an apparent movement in order to confirm the original result. As an interaction between a jet and an interstellar medium, thermal emission added to synchrotron emission is expected. Therefore, I use archival data with *Chandra* and analyze summed energy spectra for knot E and knot F.

## 2 Image analysis

### 2.1Data selection

I make a merged image with archival data from 2000 to 2018 in order to confirm the original work. Plenty of archival data are used in order to detect a faint feature. *Chandra* have a highest angular resolution of 0.5 arc second among X-ray satellites which can resolve the jet, nucleus, HST-1, knot D, knot E, knot F and knot A. The detector used is CCD. Image is taken from 2 dimension arrays of CCD and an Xray energy spectra is taken from a deposit energy in CCD. Data with a frame time of 3.1 sec is saturated for bright knots (nucleus, HST-1, knot D and knot A). A pile up distort an image. An image have to be made with no saturation. Therefore, data with a frame time of 0.4 sec is used. I use CIAO 4.15 and caldb 4.10.2.

I select data with following three steps. At step1, I check an image on a log scale by ds9 tool and remove data with a visible pile up line originated from nucleus or HST-1. At step2, I select data which nucleus and HST-1 are discriminated apparently. At step3, I check a pile up in count rate. A pile up information is made by a pileup_map command from a fits data and the count/frame is calculated for a given region file by a dmstat command. I show a position of each knot in table 1. I make region files which show a position and a radius of an extracted region from a fits data by a ds9 tool for nucleus and HST-1, knot D and A, respectively. The radius of a region is 0.5 arc second for nucleus, 0.6 arc second for HST-1, 0.75 arc second for knot D and 1 arc second for knot A. I show the count/frame from 2000 to 2018 for each knot in figure 1. The count / frame = 0.1, 0.2 corresponds to 5 %,10% pile up fraction, respectively. For knot D and knot A, all data is less then 5% pile up contamination. For nucleus and HST-1, data set from 2000 to 2012 is piled up heavily. I show an archive list with a selection from 2000 to 2018 in table 2. The observation list used by original work is also shown in table 2. I do not include 4 observations of original work in table 2, obs ID 2702 with a frame time of 3.2 sec and obs ID 7212, 5826 and 5827 with frame time of 3.1 sec. These data are piled up for bright knots. Apparent movements are occurred by a jet velocity and a precession. The apparent jet velocity for HST-1 is 6.6 c (Thimmapaa et al. 2024). This corresponds to 0".26 in 10 years. The precession for M87 is $\pm 7$ deg with a period of 11 years(Cui et al. 2023). At the position of knot A, the period is 0.6 years which difference is 0".5 (angular resolution of *Chandra*). Therefore, I use archival data within 0.6 years for one merged image. The original work did not consider a precession. I check a difference in a position between the radio core and the X ray nucleus for data which pass a pileup selection. The position of the radio core is (RA, DEC)=($12^h30^m49^s$.423, $12^{o}23'$ 28".04) (Massaro et al. 2013). A maximum difference is 0".7. I do not include data which difference is above 0".5 in order to obtain a correct image. I make three data set, one is good in both a pile up fraction and statistics, the others are poor in both a pile up fraction and statistics, as shown in table 3.

## 2.2 Merging image

At First, I use a reproject_obs command for reprojected data from fits data. A tangent point is fixed to the position of the radio core. A WCS offset is not done. This is same with an original analysis which (RA_NOM, DEC_NOM) in event file is rewrite to the radio core and no WCS offset is done. Next, I use flux_obs command for the corrected image with both an effective area and an exposure time from reprojected data. CCD has a sensitivity from 0.2 keV to 10 keV. There is a quantum efficiency degradation by a contamination of an optical filter. An energy range above 0.3 keV, especially above 0.5 keV is well calibrated. Therefore, a lower limit of an energy in image is set as 0.5 keV. I set two kinds of energy band at 0.5~2 keV and 2~7 keV, a pixel size as 0.5 arc second when flux_obs command is done. The exposure map is evaluated at 2 keV.

## 2.3 Result

The smoothing by a csmooth tool which they used is not recommended. Background calculated by a csoomth tool is taken around a spot. Background around a spot is complicated. As a background, hot gas of cluster and emission of a jet is different by a spot. The csmooth apply raw count map. Therefore, I do not recommend csmooth.

I smooth the corrected merged image with a gaussian (sigma 1, radius 2). I show the corrected merged image at energy range of 0.5~2 keV and 2~7 keV on log scale in figure 2, respectively. At 0.5~2 keV, there is dip in north region outside the jet between knot E and knot F. At 2~7 keV, there is no dip in same region. The count for a given region is obtained by a dmstat command from raw count image. The region file which show a position and a radius of an extracted region is made by a ds9 tool from the corrected image. Hot gas of Virgo cluster depends on the distance from a center(Bohringer et al. 2001). A significance is given by $(Non - Noff)/\sqrt{(Non)}$. Here, Noff is a count in the soft Xray dip region and Non is a count in a south region with same area and almost same distance from nucleus. I show a comparison of counts in table 4. The count is modified by a number of good pixels between the soft X ray dip region and the south region. At 0.5~2 keV, a significance of the soft X ray dip region is 12 sigma for data1. At 2~7 keV, same regions with 0.5~2 keV are compared and a significance of the soft X ray dip region is 1 sigma for data1. Therefore, at 2~7 keV, there is no dip. At 0.5~2 keV, the significance of soft X ray dip region is 6 sigma and 7 sigma for data2 and data3, respectively. I confirm an original result.

This is a result of an interaction between the jet and an interstellar medium for M87. A jet compress an interstellar medium and a compressed interstellar medium outside the jet absorb a soft X ray. A compressed interstellar medium is expected as molecular cloud. Molecular cloud have been detected in 40” south east of the nucleus by ALMA(Simionescu et al. 2018). For nucleus with a radius of 105 pc, both CO absorption and CO emission line were detected by ALMA (Ray & Hwang 2024). The emission area include HST-1. However, Kawanaka, Nagai and Fujita(2025) analyzed same ALMA data and found a problem in data and found no CO absorption line. On the other hands, Boizelle et al(2025) analyzed ALMA data and found no CO emission line along the kpc jet, but confirmed CO absorption line around nucleus. The existance of molecular cloud around nucleus is debated yet.

## 3.Spectral analysis

The ratio of a temperature is given by $T_2/T_1 = 2\gamma(\gamma-1)M^2/(\gamma+1)^2$ for a strong shock $M >> 1$ (Longair 1992). Here, $T_1$ is a temperature outside a jet and $T_2$ is that inside a jet. $M$ is a mach number of a shock wave. $\gamma$ is a ratio of a specific heat. When $\gamma$ is given as 5/3 for monatomic gas, $T_2/T_1 = 0.3\, M^2 >> 1$. It is possible that the gas in the jet is heated by a shock and it is observed as thermal emission. X ray energy spectra for knot E and knot F is expected as thermal emission added to synchrotron emission.

### 3.1 Data selection

For a spectral analysis, large exposure time is needed. Because poor statistics leads that any model is acceptable and a subtraction of a background is insufficient. I made the summed energy spectra with archival data in order to search for a faint feature by *Chandra*. The detector used is CCD. Emission from knot E and knot F is faint in frame time of 0.4 sec. I use archival data with a frame time of 3.1 sec. I use CIAO 4.15 and caldb 4.10.2. I select data with following two steps. With a frame time of 3.1 sec, bright spots in a jet are piled up. At step 1, I check an visible pile up line originated from nucleus or HST-1 on a log scale in the image by a ds9 tool. If a line of pile up is overlapped with a source region or a background region for knot E and knot F, correct energy spectra cannot be made. There is a line of a pile up originated from nucleus or HST-1 for obs ID 5827. However, a line of a pile up is vertical to an axis of the jet. Therefore, I also use data of obs ID 5827. At step 2, I check a pile up in count rate. A pile up information is made by pileup_map command from a fits data. The count/frame is calculated for each region file by a dmstat command. I make a region file which show a position and a radius of an extracted region from a fits data by a ds9 tool. The position of knot E and knot F for obs ID 5827 is shown in table 1. The radius is 1 arc second. I show the count/frame for knot E and knot F in figure 1. The count / frame = 0.1 corresponds to 5% pile up fraction. The observed count/frame for all data of knot E and knot F is less than a pile up 5% contamination. I show an observation list in table 5. I also make a region file which show a position and a radius of an extracted region from a fits data by ds9 tool for background. Background for the M87 jet is Cosmic Xray background, detector background, galactic emission, solar background and hot gas of cluster. Cosmic X-ray background, detector background, galactic emission and solar background can be removed by taking background region with same observation. However, hot gas of cluster is dependent on a radius from the center of Virgo cluster (Bohringer et al 2001). Therefore, a background region is neighbour to source region with same radius and almost same distance from nucleus. There is the soft X ray dip on a north region between knot E and knot F as discussed in section 2. Therefore, background is taken on only south region for knot E and knot F as shown in figure 3. The energy spectra of source region and background region are made by a specextract command from a fits data and each region file, respectively. The effective area (arf) and the energy response file(rmf) are made by a specextract command. The arf is made with no weight of an exposure time and with a correction by PSF which are suitable conditions for a point source analysis. The response (rsp) is made from an arf and a rmf for each observation by a marfrmf command. An energy spectra for a source and a background are summed by a mathpha command, respectively. The rsp for summed energy spectra is weighted with an exposure time of each observation by a addrmf command. In a summed energy spectra, data points are binned as count/bin is above 15 by a grppha command. The summed energy spectra are described with no line. The hot gas of cluster is eliminated successfully.

### 3.2 Spectra fitting

CCD has a sensitivity from 0.2 keV to 10 keV. There is a quantum efficiency degradation by a contamination of an optical filter. An energy range above 0.3 keV, especially above 0.5 keV is well calibrated. Therefore, a lower limit of an energy in energy spectra is set as 0.5 keV. XSPEC tool is used for an energy spectra analysis. At first, a power law is tried as synchrotron emission of accelerated electrons. There is a soft X ray absorption by a photo electric effect of the neutral material from M87 to earth. The column density by the 21 cm radio observation is $1.27 \mathrm{x} 10^{-20} \mathrm{cm}^{-2}$ (Bekhti et al 2016). An absorption is applied to a power law. A phabs model is used for an absorption. At first, a column density of absorption is set free. Fitting parameters are shown in table 6. For knot E, when an energy spectra are fitted with a power law, a chance probability is 0.0657. For knot F, when an energy spectra is fitted with a power law, a column density of an absorption is lower than the 21 cm observation value. Therefore, a column density is fixed to the 21 cm observation value and fitted with a power law. Fitting parameters are shown in table 7. A chance probability is 0.141. Therefore, there is no need of thermal emission.

## 4. Conclusition

I make X ray merged image of the M87 jet with archival data of *Chandra* in order to confirm original result (Dainotti et al. 2012). I select data set by a removal of pile up event completely and a consideration of a precession. I confirm the soft X ray dip outside the jet between knot E and knot F. This is an interaction between a jet and an interstellar medium. An interstellar medium is compressed by a jet and the soft X ray may be absorbed in cold cloud. Thermal emission heated by a shock is expected inside a jet. I make X ray energy spectra for knot E and knot F with archival data of *Chandra*. However, energy spectra are well described with only synchrotron emission.

## Acknowledgement

I thank *Chandra* archival data center and *Chandra* software team and CXC Help desk for a kind support.

Table 1 The position of nucleus, HST-1, knot D and knot A in obsID 18232 and knot E and knot F in obsID 5826. 1" corresponds to 78 pc

| | (RA, DEC)(J2000) | distance from nucleus | radius of region |
|---|---|---|---|
| nucleus | (12$^h$30$^m$49$^s$.46, 12$^o$23'28".0) | | 0".5 |
| HST-1 | (12$^h$30$^m$49$^s$.38, 12$^o$23'28".4) | 1".2(94 pc) | 0".6 |
| D | (12$^h$30$^m$49$^s$.27, 12$^o$23'28".9) | 2".9(226 pc) | 0".75 |
| E | (12$^h$30$^m$49$^s$.03, 12$^o$23'30".2) | 6".8(530 pc) | 1".0 |
| F | (12$^h$30$^m$48$^s$.87, 12$^o$23'31".0) | 9".4(733 pc) | 1".0 |
| A | (12$^h$30$^m$48$^s$.65, 12$^o$23'32".3) | 12".8(998pc) | 1".0 |

Table2.1 list of archival data with a frame time of 0.4 sec from 2000 to 2018.

| obs ID | obs date | no pile up in image | discrimination between nucleus and HST-1 | pile up fraction 5% in count rate | pile up fraction 10% in count rate | pile up fraction 5% in count rate | pile up fraction 10% in count rate | distance of nucleus from radio core | used by original work |
|---|---|---|---|---|---|---|---|---|---|
| | | | | nucleus | nucleus | HST-1 | HST-1 | | |
| 1808 | 2000/7/30 | o | x | | | | | | o |
| 3084 | 2002/2/12 | o | o | x | o | x | o | | o |
| 3085 | 2002/1/16 | o | o | x | x | x | x | | o |
| 3086 | 2002/3/30 | o | o | x | x | x | o | | o |
| 3087 | 2002/6/8 | o | o | x | x | x | x | | o |
| 3088 | 2002/7/24 | o | o | x | x | x | x | | o |
| 3975 | 2002/11/17 | o | o | x | x | x | x | | o |
| 3976 | 2002/12/29 | o | o | x | o | x | x | | o |
| 3977 | 2003/2/4 | o | o | x | x | x | x | | o |
| 3978 | 2003/3/9 | o | o | x | x | x | x | | o |
| 3979 | 2003/4/14 | o | o | x | x | x | x | | o |
| 3980 | 2003/5/18 | o | o | x | x | x | x | | o |
| 3981 | 2003/7/3 | o | x | | | | | | |
| 3982 | 2003/8/8 | o | x | | | | | | |
| 4917 | 2003/11/11 | o | x | | | | | | |
| 4918 | 2003/12/29 | o | x | | | | | | |
| 4919 | 2004/2/12 | o | x | | | | | | |
| 4921 | | x | x | | | | | | |
| 4922 | | x | x | | | | | | |
| 4923 | | x | x | | | | | | |
| 5737 | | x | o | x | x | x | x | | |
| 5738 | | x | x | | | | | | |
| 5739 | | x | o | x | x | x | x | | |
| 5740 | | x | x | | | | | | |
| 5741 | | x | x | | | | | | |
| 5742 | | x | x | | | | | | |
| 5743 | | x | x | | | | | | |
| 5744 | | x | x | | | | | | |
| 5745 | | x | x | | | | | | |
| 5746 | | x | x | | | | | | |
| 5747 | | x | x | | | | | | |
| 5748 | | x | x | | | | | | |
| 6299 | | x | o | x | x | x | x | | |
| 6300 | | x | o | x | x | x | x | | |
| 6301 | 2006/2/19 | o | o | x | x | x | x | | |
| 6302 | 2006/3/30 | o | x | | | | | | |
| 6303 | 2006/5/21 | o | o | x | o | x | x | | |

Table 2.2 continued

| obs ID | obs date | no pile up in image | discrimination between nucleus and HST-1 | pile up fraction 5% in count rate | pile up fraction 10% in count rate | pile up fraction 5% in count rate | pile up fraction 10% in count rate | distance of nucleus from radio core | used by original work |
|---|---|---|---|---|---|---|---|---|---|
| | | | | nucleus | nucleus | HST-1 | HST-1 | | |
| 6304 | | x | o | x | x | x | x | | |
| 6305 | 2006/8/2 | o | o | x | x | x | x | | |
| 7348 | | x | x | | | | | | |
| 7349 | | x | x | | | | | | |
| 7350 | | x | x | | | | | | o |
| 7351 | 2007/3/24 | o | x | | | | | | o |
| 7352 | 2007/5/15 | o | x | | | | | | o |
| 7353 | | x | x | | | | | | o |
| 7354 | | x | x | | | | | | o |
| 8510 | 2007/2/15 | o | x | | | | | | o |
| 8511 | 2007/2/18 | o | x | | | | | | o |
| 8512 | 2007/2/21 | o | x | | | | | | o |
| 8513 | 2007/2/24 | o | o | x | x | x | x | | o |
| 8514 | 2007/3/12 | o | x | | | | | | o |
| 8515 | 2007/3/14 | o | x | | | | | | o |
| 8516 | 2007/3/19 | o | o | x | x | x | x | | o |
| 8517 | 2007/3/22 | o | o | x | x | x | x | | o |
| 8575 | 2007/11/25 | o | o | x | x | x | x | | o |
| 8576 | 2008/1/4 | o | o | x | x | x | x | | o |
| 8577 | 2008/2/16 | o | o | x | x | x | x | | o |
| 8578 | 2008/4/1 | o | o | x | x | x | x | | o |
| 8579 | 2008/5/15 | o | o | x | x | x | x | | o |
| 8580 | 2008/6/24 | o | x | | | | | | o |
| 8581 | 2008/8/7 | o | o | x | o | x | x | | o |
| 10282 | 2008/11/17 | o | o | x | o | x | o | 0.1" | o |
| 10283 | 2009/1/7 | o | o | x | o | x | o | 0.2" | o |
| 10284 | 2009/2/20 | o | o | x | o | x | o | 0.1" | o |
| 10285 | 2009/4/1 | o | o | x | o | x | o | 0.2" | o |
| 10286 | 2009/5/13 | o | o | x | o | x | o | 0.1" | o |
| 10287 | 2009/6/22 | o | o | x | o | x | o | 0.2" | o |
| 10288 | 2009/12/15 | o | o | | x | | o | | |
| 11512 | 2010/4/11 | o | o | | x | | o | | |
| 11513 | 2010/4/13 | o | o | x | o | x | o | 0.1" | |
| 11514 | 2010/4/15 | o | o | x | o | x | o | 0.2" | |
| 11515 | 2010/4/17 | o | o | x | o | x | o | 0.1" | |
| 11516 | 2010/4/20 | o | o | x | o | x | o | 0.4" | |
| 11517 | 2010/5/5 | o | o | x | o | x | o | 0.5" | |

Table 2.3 continued

| obs ID | obs date | no pile up in image | discrimination between nucleus and HST-1 | pile up fraction 5% in count rate | pile up fraction 10% in count rate | pile up fraction 5% in count rate | pile up fraction 10% in count rate | distance of nucleus from radio core | used by original work |
|---|---|---|---|---|---|---|---|---|---|
| | | | | nucleus | nucleus | HST-1 | HST-1 | | |
| 11518 | 2010/5/9 | o | o | x | o | o | o | 0.5" | |
| 11519 | 2010/5/11 | o | o | x | o | x | o | 0.1" | |
| 11520 | 2010/5/14 | o | o | x | o | x | o | 0.3" | |
| 13964 | 2011/12/5 | o | o | x | o | o | o | 0.6" | |
| 13965 | 2012/2/25 | o | x | | | | | | |
| 14973 | 2013/3/12 | o | o | x | o | o | o | 0.3" | |
| 14974 | 2012/12/12 | o | o | x | o | o | o | 0.1" | |
| 16042 | 2013/12/26 | o | o | o | o | o | o | 0.2" | |
| 16043 | 2014/4/2 | o | x | | | | | | |
| 17056 | 2014/12/17 | o | o | o | o | o | o | 0.2" | |
| 17057 | 2015/3/19 | o | o | o | o | o | o | 0.2" | |
| 18232 | 2016/4/27 | o | o | o | o | o | o | 0.5" | |
| 18233 | 2016/2/23 | o | o | o | o | o | o | 0.6" | |
| 18781 | 2016/2/24 | o | o | o | o | o | o | 0.3" | |
| 18782 | 2016/2/26 | o | o | o | o | o | o | 0.4" | |
| 18783 | 2016/4/20 | o | o | o | o | o | o | 0.3" | |
| 18809 | 2016/3/12 | o | o | o | o | o | o | 0.7" | |
| 18810 | 2016/3/13 | o | o | o | o | o | o | 0.1" | |
| 18811 | 2016/3/14 | o | o | o | o | o | o | 0.4" | |
| 18812 | 2016/3/16 | o | o | o | o | o | o | 0.3" | |
| 18813 | 2016/3/17 | o | o | o | o | o | o | 0.5" | |
| 18836 | 2016/4/28 | o | o | o | o | o | o | 0.5" | |
| 18837 | 2016/4/30 | o | o | o | o | o | o | 0.5" | |
| 18838 | 2016/5/28 | o | o | o | o | o | o | 0.7" | |
| 18856 | 2016/6/12 | o | o | o | o | o | o | 0.5" | |
| 19457 | 2017/2/15 | o | o | o | o | o | o | 0.5" | |
| 19458 | 2017/2/16 | o | o | o | o | o | o | 0.3" | |
| 20034 | 2017/4/11 | o | o | o | o | o | o | 0.4" | |
| 20035 | 2017/4/14 | o | x | | | | | | |
| 20488 | 2018/1/4 | o | x | | | | | | |
| 20489 | 2018/3/21 | o | x | | | | | | |
| 21075 | 2018/4/22 | o | x | | | | | | |
| 21076 | 2018/4/24 | o | x | | | | | | |

Table 3 The data set for merged image in this paper

| label | pile up fraction | observation date | obs ID | exposure time |
|---|---|---|---|---|
| data1 | 5% | 2016 Feb~2016 Jun | 18810~18813, 18232, 18781~18783, 18836, 18837, 18856 | 224 ks |
| data2 | 10% | 2008 Nov~2009 Jun | 10282~10287 | 28 ks |
| data3 | 10% | 2010 Apr~2010 May | 11513~11520 | 37 ks |

Table 4 The count for soft Xray dip region and south region for each data set.

| | | count in soft Xray dip region (number of good pixel) | count in south region (number of good pixel) |
|---|---|---|---|
| data1 | 0.5~2 keV | 1188(26) | 1602(25) |
| | 2~7 keV | 170(26) | 177(25) |
| data2 | 0.5~2 keV | 228(26) | 319(25) |
| | 2~7 keV | 17(26) | 30(25) |
| data3 | 0.5~2 keV | 279(27) | 393(25) |
| | 2~7 keV | 34(27) | 31(25) |

Table 5 The observation log used for spectra analysis. Data with a frame time of 3.2 sec is used.

| obs ID | PI | obs date |
|---|---|---|
| 5826 | Forman | 2005.3 |
| 5827 | Forman | 2005.5 |
| 5828 | Forman | 2005.11 |
| 7210 | Forman | 2005.11 |
| 7211 | Forman | 2005.11 |
| 7212 | Forman | 2005.11 |
| total exposure time | | 480 ks |

Table 6 Fitting parameters with a power law for summed energy spectra. The column density of an absorption is set free. An error is 90% confidence level statistical error.

| | E | F |
|---|---|---|
| $N_H$(x$10^{20}$ cm$^{-2}$) | $3.78^{+3.08}_{-2.96}$ | 0.00 |
| photon index | $2.36^{+0.10}_{-0.10}$ | $2.59^{+0.08}_{-0.09}$ |
| 1keV flux(ph/cm$^2$/s/keV) | $3.99^{+0.39}_{-0.36}$ (x$10^{-5}$) | $2.09^{+0.10}_{-0.10}$ (x$10^{-5}$) |
| $\chi^2$/d.o.f (d.of) | 1.165(176) | 1.122(137) |

Table 7 Fitting parameters with a power law for summed energy spectra of knot F. Column density of an absorption is fixed to the 21cm observation value. An error is 90% confidence level statistical error.

| photon index | $2.63^{+0.09}_{-0.09}$ |
|---|---|
| 1keV flux(ph/cm$^2$/s/keV) | $2.17^{+0.11}_{-0.10}$ (x$10^{-5}$) |
| $\chi^2$/d.o.f (d.of) | 1.130(138) |

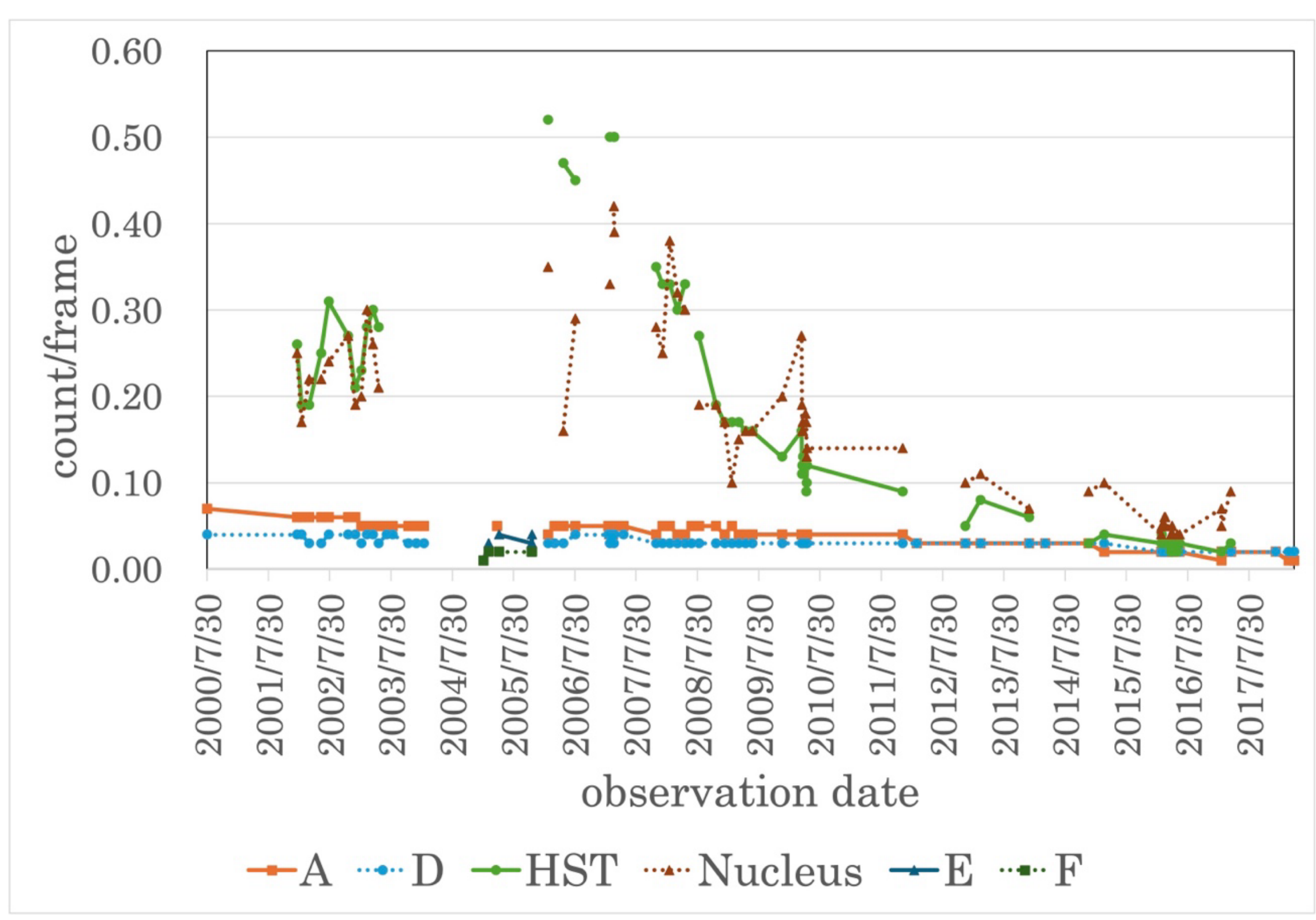

Figure 1 The count / frame from 2000 to 2018 for each knot. For nucleus, HST-1, kont D and knot A, frame time is 0.4 sec. For knot E and knot F, frame time is 3.2 sec.

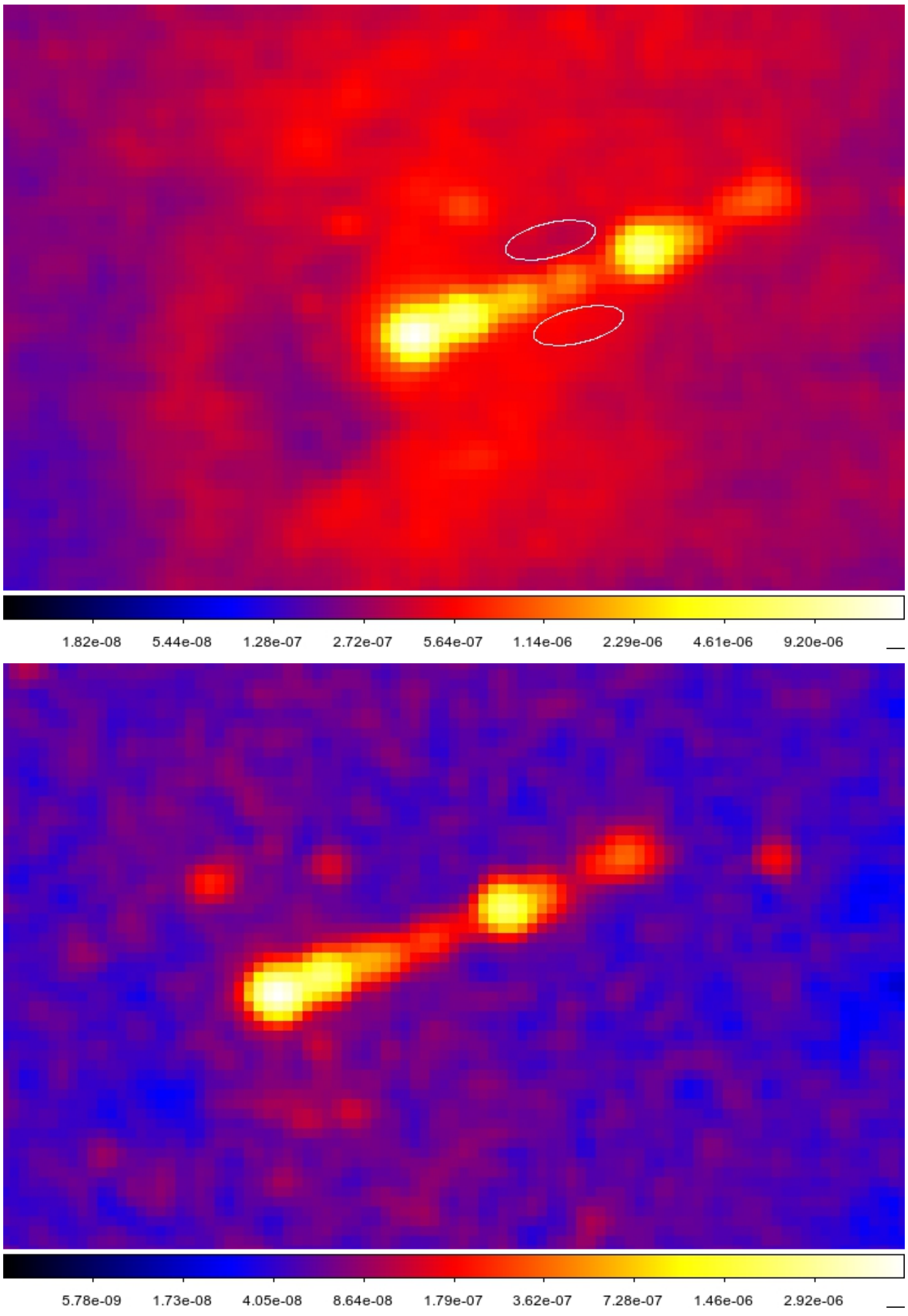

Figure 2 The corrected merged image of the M87 jet at 0.5~2 keV(top) and 2~7 keV(bottom) for data1. The smooth by gaussian is done. At 0.5~2 keV, the soft X ray dip is shown in the north outside between knot E and knot F by white circle and the south region for a comparison is shown by white circle

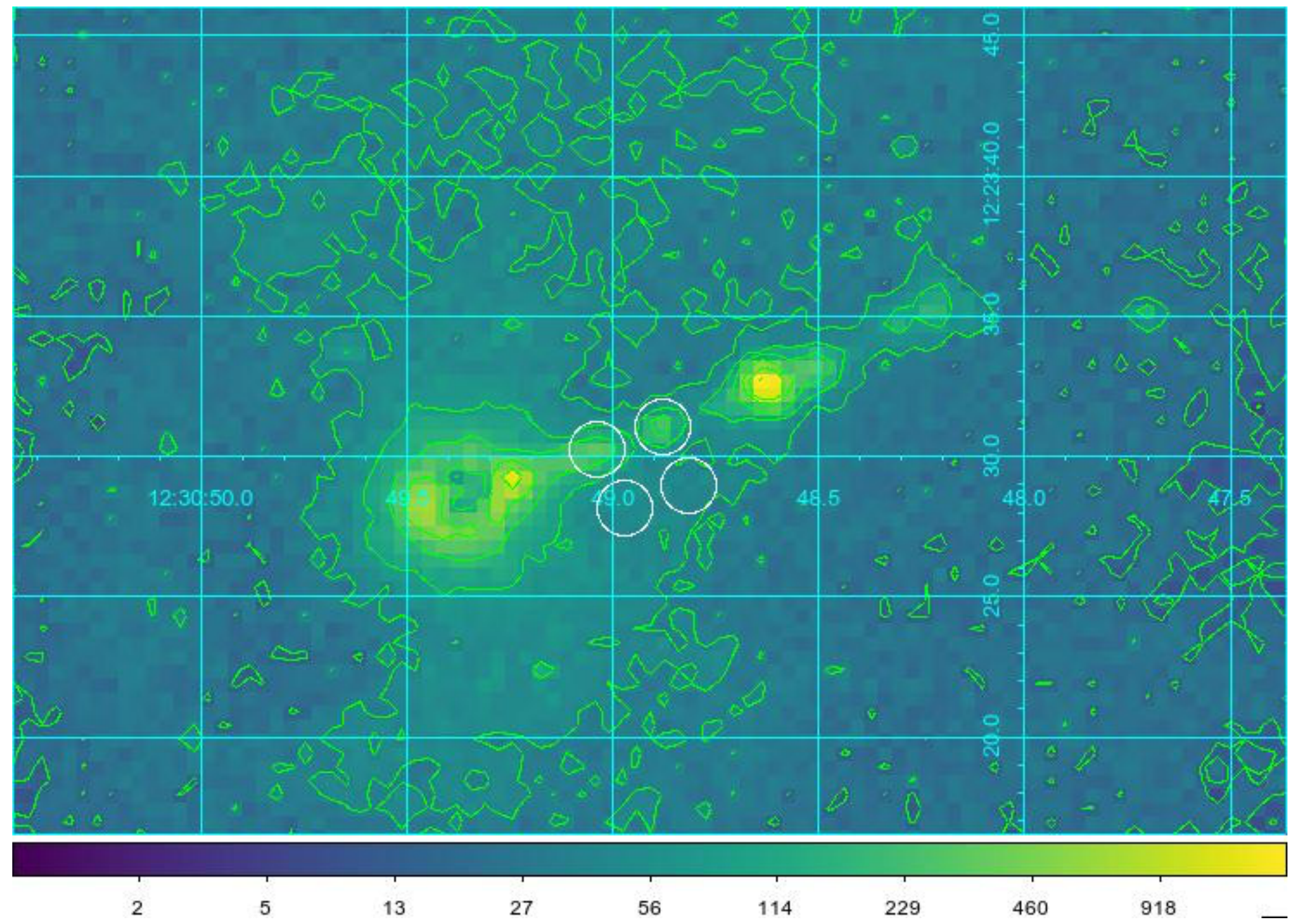

Figure 3 The smoothed image of M87 jet for obs ID 5826. A source region and background region for knot E and knot F are shown by white circle from left